\documentstyle[aps,epsf,epsfig,prl,psfig]{revtex}
\newcommand{\avg}[1]{\langle{#1}\rangle}

\newcommand{\be}{\begin{equation}\FL}
\newcommand{\ee}{\end{equation}}
\newcommand{\beas}{\begin{eqnarray*}}
\newcommand{\eeas}{\end{eqnarray*}}
\newcommand{\bea}{\begin{eqnarray}\FL}
\newcommand{\eea}{\end{eqnarray}}
\newcommand{\req}[1]{(\ref{#1})}
\def\sign{\hbox{sign}\,}

\begin{document}

\twocolumn[\hsize\textwidth\columnwidth\hsize\csname
@twocolumnfalse\endcsname
\title{Statistical mechanics of systems with heterogeneous agents: 
Minority Games}
\author{Damien Challet$^{(1)}$, Matteo Marsili$^{(2)}$ 
and Riccardo Zecchina$^{(3)}$}
\address{$^{(1)}$Institut de Physique Th\'eorique, 
Universit\'e de Fribourg, CH-1700\\
$^{(2)}$Istituto Nazionale per la Fisica della Materia (INFM),
Trieste-SISSA Unit, V. Beirut 2-4, Trieste I-34014,\\
$^{(3)}$The Abdus Salam International Centre for Theoretical Physics
Strada Costiera 11, P.O. Box 586, I-34100 Trieste}
\date{\today}
\maketitle
\widetext

\begin{abstract}

We study analytically a simple game theoretical model of heterogeneous
interacting agents. 
We show that the stationary state of the system
is described by the ground state of a disordered spin model which
is exactly solvable within the simple replica symmetric ansatz.
Such a stationary state differs from the Nash equilibrium 
where each agent maximizes her own utility. The latter
turns out to be characterized by a replica symmetry broken
structure.
Numerical results fully agree with our analytic findings.
\end{abstract}

]
\narrowtext

Statistical mechanics of disordered systems 
provides analytical and numerical tools for the description 
of complex systems, which have found applications in many
interdisciplinary areas\cite{MPV}. 
When the precise realization of the interactions in
an heterogeneous system
is expected not to be crucial for the overall macroscopic behavior, then
the system itself can be modeled as having random interactions 
drawn from an appropriate distribution. Such an approach 
appears to be very promising 
also for the study of systems with many heterogeneous agents, such as  
markets, which have recently attracted much interest in the 
statistical physics community \cite{pw,CZ1}. Indeed it provides a 
workable alternative to the so called {\em representative agent} 
approach of micro-economic theory, where assuming that agents
are identical, one is lead to a theory with one single 
(representative) agent\cite{micro}.

In this Letter we present analytical results for a simple 
model of heterogeneous interacting agents, the so called minority
game (MG)\cite{CZ1,MGrev}, which is a toy model of $N$
agents interacting through a global quantity representing a 
market mechanism. 
Agents aim at anticipating market movements by following a simple adaptive
dynamics inspired at Arthur's {\em inductive reasoning}\cite{Arthur}. 
This is based on simple {\em speculative}
strategies that take advantage of the available public information 
concerning the recent market history, which can take the form of one of
$P$ patterns. Numerical studies \cite{CZ1,Savit,cm,cavagna} have
shown that the model displays a remarkably rich behavior. 
The relevant control parameter\cite{CZ1,Savit}  turns out to be the ratio
$\alpha=P/N$ between the ``complexity'' of information $P$ and the number
$N$  of agents, and the model undergoes a phase transition with symmetry 
breaking\cite{cm} independently of the  origin of information\cite{cavagna}.

We shall limit the discussion on the interpretation of the model --
which is discussed at some length in refs. \cite{CZ1,Savit} -- to a 
minimum and rather focus on its mathematical structure and to the analysis 
of its statistical properties for $N\gg 1$. 
Our main aim is indeed to show that the model can be
analyzed within the framework of statistical mechanics of disordered
system\cite{MPV}.

We find that dynamical steady states can be mapped onto the ground state
properties of a model very similar to that proposed in ref.\cite{bz}
in the context of optimal dynamics for attractor neural networks.
There\cite{bz} one shows that the minimization of the interference noise
is equivalent to maximizing the dynamical stability of each device composing 
the system. Conversely, we show that 
the individual utility maximization in interacting agents systems is
equivalent to the minimization of a global function.
We also find that different learning models lead to different 
patterns of replica symmetry breaking.

The model is defined as follows\cite{cm}:
Agents live in a world which can be in one of $P$ {\em states}. 
These are labelled by an integer $\mu=1,\ldots,P$ which encodes
all the information available to agents. For the moment being, we follow
ref. \cite{cavagna} and assume that this information concerns some external
system so that $\mu$ is drawn from a uniform distribution
$\varrho^\mu=1/P$ in $\{1,\ldots,P\}$. 
Each agent $i=1,\ldots,N$ can choose between one of two 
{\em strategies}, labeled by a spin variable $s_i\in\{\pm 1\}$,
which prescribes an {\em action} $a_{s_i,i}^\mu$ for each state $\mu$.
Strategies may be ``look up tables'', behavioral rules \cite{CZ1,Arthur} or 
information processing devices.
The actions $a_{s,i}^\mu$ are drawn from a 
bimodal distribution $P(a_{s,i}^\mu=\pm 1)=1/2$ for all 
$i,s$ and $\mu$ and they will play the role of quenched disorder\cite{MPV}.
Hence there are only two possible actions -- such as ``do something'' 
($a_{s,i}^\mu=1$) or ``do the opposite'' ($a_{s,i}^\mu=-1$).
It is convenient\cite{cm} to make the dependence on $s$ explicit
in $a_{s,i}^\mu$, introducing $\omega_i^\mu$ and $\xi_i^\mu$ so that
$a_{s,i}^\mu=\omega_i^\mu+ s\xi_i^\mu$\cite{nota}. If agent $i$ chooses 
strategy $s_i$ and her opponents choose strategies $s_{-i}
\equiv \{s_j,j\ne i\}$, in state $\mu$, she receives a payoff
\be
u_i^\mu(s_i,s_{-i})=-a_{s_i,i}^\mu G(A^\mu),
\label{ui}
\ee
where, defining $\Omega^\mu=\sum_j\omega_j^\mu$, 
\be
A^\mu=\sum_j a_{s_j,j}^\mu=\Omega^\mu +\sum_j\xi_j^\mu s_j.
\ee 
The function $G(x)$, which describes the market mechanism, is such that
$x\,G(x)>0$ for all $x$ so that the total payoff to agents is always
negative: the majority of agents receives a negative payoff whereas only the
minority of them gain. Note that the agent--agent interaction, which comes from
the aggregate quantity $G(A^\mu)$, is of mean-field character.

The game defined by the payoffs in Eq. \req{ui} can be analyzed along the
lines of game theory\cite{games} by looking for its Nash equilibria in 
the strategies space $\{s_j, j=1,\ldots,N\}$. Before doing this, we prefer
to discuss the dynamics of {\em inductive agents} following refs. 
\cite{CZ1,Savit,cm}: There, the game is repeated many times and agents try 
to estimate empirically which of the two strategies they have is the best 
one, using past observations. More precisely, each agent $i$ assigns a 
{\em score} $U_{s,i}(t)$ to her $s^{\rm th}$ strategy at time $t$, 
and we assume, as in ref. \cite{cavagna2}, that she chooses that
strategy with probability\cite{logit}
\be
\pi_{s,i}(t)\equiv {\rm Prob}\{s_i(t)=s\}=C e^{\Gamma U_{s,i}(t)}
\label{sit}
\ee
with $C^{-1}=\sum_{s'}e^{\Gamma U_{s',i}(t)}$ and $\Gamma>0$.
The scores are initially set to $U_{s,i}(0)=0$, and they are updated as
\be
U_{s,i}(t+1)=U_{s,i}(t)-a_{s,i}^{\mu(t)}G(A^{\mu(t)})/P.
\label{uit}
\ee
The idea is that if a strategy $s$ has predicted the right sign
i.e. if $a_{s,i}^\mu=-{\rm sign}\,G(A^\mu)$, its score, and hence its
probability of being used, increases. Note that 
$a_{s,i}^{\mu}G(A^{\mu})$ in Eq. \req{uit} is {\em not} the payoff 
$u^\mu_i(s,s_{-i})$ which agent $i$ would have received if she had 
actually played strategy $s\ne s_i(t)$. 
Indeed $G(A^\mu)$ depends on the strategy
$s_i(t)$ that agent $i$ has actually played through $A^\mu$. 
Agents in the MG neglect this effect and behave as if they were facing 
an external process $G(A^\mu)$ rather than playing against other $N-1$ 
agents. This may seem reasonable for $N\gg 1$ since the relative
dependence of aggregate quantities on  each agent's choice is expected to be
small. We shall see below (see Eq. \ref{uitilde}) that this is not true:
if agents consider the impact of their actions on $A^\mu$, the
collective behavior changes considerably. 

We focus on the linear case $G(x)=x$, which allows for a simple
treatment. Other choices, such as the original one $G(x)=\sign x$, lead
to similar conclusions, as it will be discussed
elsewhere\cite{forthcoming}. With this choice, the total losses of 
agents is $-\sum_i u_i^\mu=(A^\mu)^2$. The time average $\sigma^2$ of
$(A^\mu)^2$ is shown in Fig. 1, as a function of $\alpha\equiv P/N$.  The
system shows a complex behavior characterized, among other things, by a phase
transition at $\alpha_c\simeq 0.34$\cite{cm} where $\sigma^2$ shows a cusp and
a small $\alpha$ phase where $\sigma^2$ increases with $\Gamma$\cite{cavagna2}.

In order to uncover this behavior, let us focus on the long time behavior
of the dynamics. The key observation is that, in the long run, the score
of a strategy depends on its performance in all $P$ states. Hence, the
behavior of agents will change systematically only on time-scales of order 
$P$. This suggests to introduce the rescaled time $\tau=t/P$. As $P\to \infty$,
any finite interval $d\tau=\Delta t/P$ is made of infinitely many time steps
and we can use the law of large numbers to approximate time averages with 
statistical averages over the variables $\mu(t)$ and $s_i(t)$ from their 
respective distributions $\varrho^\mu$ and $\pi_{s,i}$. 
We henceforth use the notation $\overline{o}=\sum_\mu\varrho^\mu o^\mu$
for averages over $\mu$ and $\avg{\cdot}$ for averages on $s_i(t)$ and
we define $m_i(\tau)\equiv \avg{s_i(t)}$. With this notations,
$\sigma^2$ reads:
\be
\sigma^2\!=\overline{\avg{A^2}}=
\overline{\Omega^2}\!+\!\sum_i\left[\overline{\xi_i^2}\!+\!
2\overline{\Omega\xi_i}m_i\right]\!+\!
\sum_{i\neq j} \overline{\xi_i\xi_j}m_im_j
\label{s2}
\ee
where we have used statistical independence of $s_i$, i.e.
$\avg{s_i s_j}=m_i m_j+(1-m_i^2)\delta_{i,j}$.
The evolution of scores $U_{s,i}$ in continuum time $\tau$,
is obtained iterating Eq. \req{uit} for $\Delta t=Pd\tau$ time steps.
Using Eq. \req{sit} in the form $m_i=\tanh[\Gamma (U_{+1,i}-U_{-1,i})]$,
we find
\be
\frac{dm_i}{d\tau}=- 2 \Gamma (1-m_i^2)\left[\overline{\Omega\xi_i}
+\sum_j\overline{\xi_i\xi_j}m_j\right].
\label{dynID}
\ee
This can be easily written as a gradient descent dynamics 
$\frac{dm_i}{d\tau} =-\Gamma (1-m_i^2)\frac{\partial H}
{\partial m_i}$ which minimizes the Hamiltonian 
\be
H=\overline{\avg{A}^2}=
\sigma^2-\sum_i  \overline{\xi_i^2} (1-m_i^2).
\label{HID}
\ee
As a function of $m_i$, $H$ is a positive definite quadratic form,
which has a unique minimum. This implies that {\em the stationary state
of the MG is described by the ground state properties of
$H$}. It is easy to see\cite{forthcoming} that $H$ is closely
related to  the order parameter $\theta=\sqrt{\overline{\avg{ \sign A}^2}}$ 
introduced in \cite{cm}, which is a measure of the system 
predictability\cite{cm}. Indeed $H\propto
\theta^2$ when $\theta$ is small, suggesting
that inductive agents actually minimizes predictability rather than 
their collective losses $\sigma^2$. 

It is possible to study the ground state properties of $H$ in Eq.
\req{HID} using the replica method \cite{MPV}. First we introduce
an inverse temperature $\beta$\cite{notaT} and compute
the average over the disorder variables $\Xi=\{ a_{s,i}^\mu
\}$ of the partition function of $n$ replicas of the system, 
$\langle Z^n\rangle_{\Xi}$. Next we 
perform an analytic continuation for non-integer values of $n$, thus
obtaining $\langle\ln Z\rangle_{\Xi} = \lim_{n\rightarrow 0}
\frac{\langle Z^n\rangle_{\Xi} -1}{n}$. 
The `free energy' $F_{ID}=-\langle\ln Z\rangle_{\Xi}/\beta$ 
depends on the the overlap matrix $Q_{a,b}=\avg{m_{i}^a m_{i}^b}$ 
($a,b=1,...n$, $a\neq b$) and on the order parameter 
$Q_a=\frac{1}{N}\sum_i(m_i^a)^2$,
together with their Lagrange multipliers $r_{a,b}$ and $R_a$ respectively. 
$F_{ID}$ can be calculated using a saddle point method
that, within the replica symmetric (RS) ansatz 
$Q_{a,b}=q$, $r_{a,b}=r$ (for all
$a<b$), and $Q_a=Q$, $R_a=R$ (for all $a$), leads to 
\beas
F_{ID}&=&\frac{\alpha}{2}\frac{1+q}{\alpha+\beta(Q-q)}+
\frac{\alpha}{2\beta}\log\left[1+\frac{\beta(Q-q)}{\alpha}\right]\\
&+&\frac{\beta}{2}(RQ-rq)-\frac{1}{\beta}\int d\Phi(\zeta)
\log\int_{-1}^1 ds\, e^{-\beta V(s|\zeta)}
\eeas
where $V(x|\zeta)=\beta(r-R)\frac{x^2}{2}-\sqrt{r}\zeta x$ 
and $\Phi$ is the normal distribution. The ground state properties
of $H$ are obtained solving the saddle point equations\cite{MPV} in the 
limit $\beta\to\infty$.
Fig. 1 compares the analytic and numerical findings for $\sigma^2$.
For $\alpha>\alpha_c=0.33740\ldots$,
the solution leads to $Q=q<1$ and a ground state energy $H_0 >0$.
$H_0 \to 0$ as $\alpha\to\alpha_c^+$ and $H_0=0$ for 
$\alpha\le \alpha_c$. 

This confirms the conclusion $\avg{A^\mu}=0$ $\forall \mu$\cite{cm}
(or $\theta=0$) for $\alpha\le \alpha_c$ and it implies the relation
\be
\sigma^2=\sum_i \overline{\xi_i^2} (1-m_i^2)\cong \frac{N}{2}(1-Q), 
~~~~\alpha\le\alpha_c.
\ee

The RS solution is stable against
replica symmetry breaking (RSB) for any $\alpha$, as expected
from  positive definiteness of $H$.
Following ref.\cite{bz}, we compute the probability distribution of the
strategies, which for $\alpha>\alpha_c$ is bimodal and it assumes the 
particularly simple form
\begin{equation}
{\cal P}(m)= \phi(z) [\delta(m-1)+\delta(m+1)]+
\frac{z}{\sqrt{2 \pi}}e^{-(z m)^2/2}
\label{Prob}
\end{equation}
with $z=\sqrt{\alpha/(1+Q)}$ ($Q$ taking its saddle point value) and
where $\phi(z)=(1-{\rm Erf}(z/\sqrt{2}))/2$ is the fraction of 
frozen agents (those who always play one and the same strategy).
Below $\alpha_c$, ${\cal P}(m)$ is continuous, i.e. $\phi=0$ 
in agreement with numerical findings\cite{cm}.

At the transition the spin susceptibility
$\chi=\lim_{\beta\to\infty}\beta(Q-q)$ diverges as $\alpha\to\alpha_c^+$
and it remains infinite for all $\alpha\le\alpha_c$. This is because the
ground state is degenerate in many directions (zero modes) and an
infinitesimal perturbation can cause a finite shift in the equilibrium
values of $m_i$. This implies that in the long run, the dynamics
\req{dynID} leads to an equilibrium state which depends on the initial
conditions $U_{s,i}(t=0)$. The under-constrained nature of the system
is also responsible for the occurrence of anti-persistent effects for
$\alpha<\alpha_c$\cite{cm}. The periodic motion in the subspace $H=0$ is 
probably induced by inertial terms $d^2U_{s,i}/d\tau^2$ which we have 
neglected, and which require a more careful study of dynamical solutions 
of Eqs. (\ref{sit},\ref{uit}). It is however clear that the amplitude of the 
excursion of $U_{+1,i}(t)-U_{-1,i}(t)$ decreases with $\Gamma$, by the
smoothing effect of Eq. \req{sit}. When this amplitude becomes of the same 
order of $1/\Gamma$ anti-persistence is destroyed, which explains the sudden 
drop of $\sigma^2$ with $\Gamma$ found in ref. \cite{cavagna2}.

A natural question arises: is this state individually optimal, i.e. is it a 
Nash equilibrium of the game where agents maximize the expected utility
$\overline{u_i}=-\overline{a_{s,i}A}$? One way to find the Nash equilibria 
is to consider stationary solutions of the multi-population replicator 
dynamics\cite{evol}. This takes the form of an equation for the so called
{\em mixed} strategies, i.e. for the probabilities $\pi_{s,i}$ with which
agent $i$ plays strategy $s$. In terms of $m_i=\pi_{+,i}-\pi_{-,i}$, 
with a little algebra, these equations \cite{evol} read
\be
\frac{dm_i}{d\tau}=(1-m_i^2)\frac{\partial \overline{u_i}}
{\partial m_i}.
\label{dyn}
\ee
Observing that $\frac{\partial \overline{u_i}}{\partial m_i}=-
\frac{\partial \sigma^2}{\partial m_i}$, we can rewrite Eq. \req{dyn} 
as a gradient descent dynamics which minimizes a global function 
which is {\em exactly} the total loss $\sigma^2$ of agents. Nash equilibria
then correspond to the local minima of $\sigma^2$ in the domain
$[-1,1]^N$. The quadratic form $\sigma^2$ is not positive definite, which
means that there shall be many local minima and the Nash equilibrium is not
unique. It is easy to see \cite{forthcoming} that Nash equilibria 
are in {\em pure} strategies, i.e. $m_i^2=1$ $\forall i$, which implies
$\sigma^2=H$, by Eq. \req{HID}.
A detailed characterization of the Nash equilibria shall be
given elsewhere\cite{forthcoming}.
The best Nash equilibrium can be studied applying the replica method to 
$\sigma^2$ for $\beta\to\infty$. The multiplicity of Nash equilibria
(meta-stable states) manifests itself in the occurrence of replica symmetry
breaking for any $\alpha>0$ with a non-vanishing $\sigma^2/N$
\cite{forthcoming}. The simple RS solution, though incorrect, provides a
close lower bound
$F_{NE}^{(RS)}=F_{ID}+\frac{1}{2}(1-Q)$ to $\sigma^2/N$ for
$\beta\to\infty$ (see Fig. 1). For $\alpha > 1/\pi$, we  have
$Q=q=1$ and $F_{NE}^{(RS)}(\beta=\infty)=[1-1/\sqrt{\pi\alpha}]^2$ 
positive, whereas $1=Q<q$ and $F_{NE}^{(RS)}=0$ for $\alpha <1/\pi$.

Fig. 1 shows that in a Nash equilibrium agents perform way 
better than in the MG. This is the consequence of the fact that agents do 
not take into account their impact on the market (i.e. on $A^\mu$) when 
they update the scores of their strategies by Eq. \req{uit}. It is
indeed known\cite{rust} that reinforcement-learning dynamics based on 
Eq. (\ref{sit}) is closely related to the replicator dynamics
and hence it converges to rational expectation outcomes, i.e. to Nash 
equilibria. More precisely, ref. \cite{rust} suggests that this occurs if 
Eq. \req{uit} is replaced with
\be
U_{i,s}(t+1)=
U_{i,s}(t)+u_i^{\mu(t)}(s,s_{-i}(t))/P
\label{uitilde}
\ee
Now $U_{s,i}(t)$ is proportional to 
the cumulated payoff that agent $i$ would
have received had she always played strategy $s$ (with other agents playing
what they actually played) until time $t$. 
As Fig. 1 again shows this leads to results which coincide with those of 
the Nash equilibrium. It is remarkable that the (relative) difference between
Eqs. \req{uit} and \req{uitilde} is small, i.e. of order $1/A^\mu\sim
1/\sqrt{N}$. Yet, it is {\em not negligible} because, when averaged over all
states $\mu$ it produces a finite effect, specially for $\alpha<\alpha_c$
and it affects considerably the nature of the stationary state.
This term has the same origin of the cavity reaction term in spin 
glasses\cite{MPV}.
In order to follow Eq. \req{uitilde} agents need to know the payoff they would 
have received for any strategy $s$ they could have played. That may not be
realistic in complex situations where agents know only the payoffs they receive
and are unable to disentangle their contribution from $G(A^\mu)$. However
agents can account approximately for their impact on the market by adding a
cavity term $+\eta\delta_{s,s_i(t)}$ to Eq. \req{uit} which ``rewards'' the 
strategy $s_i(t)$ used with respect to those $s\ne s_i(t)$ not used. 
The most striking effect of this new term, as discussed elsewhere 
\cite{forthcoming} in detail, 
is that for $\alpha<\alpha_c$ an {\em infinitesimal}  $\eta>0$ is sufficient 
to cause RSB and to reduce $\sigma^2/N$ by a {\em finite} 
amount.

So far, the information $\mu(t)$ was randomly and independently drawn at each
time $t$ form the distribution $\varrho^\mu=1/P$. In the original version 
of the
MG\cite{CZ1} $\mu$ is instead endogenously determined by the
collective dynamics of agents: $\mu(t)$ indeed labels the sequence of the last
$M=\log_2 P$ ``minority" signs -- i.e. $\mu(t+1)=[2\mu(t)+1]_{{\rm mod}~P}$ if
$A^{\mu(t)}>0$ and $\mu(t+1)=[2\mu(t)]_{{\rm mod}~P}$ otherwise. The
idea\cite{CZ1} is that the information refers to the recent past history of the
market, and agents try to guess trends and patterns in the time evolution of the
process $G(A^{\mu(t)})$. We may say that $\mu(t)$ is {\em
endogenous} information, since it refers to the market itself, as opposed to
the {\em exogenous} information case discussed above.

Numerical simulations \cite{cavagna} show that the collective behavior of the MG
-- based on Eq. \req{uit} -- under endogenous information is the same as that
under exogenous information. Within our approach, the relevant
feature of the dynamics of $\mu(t)$ is its stationary state distribution
$\varrho^\mu$. The key point is that a finite fraction $1-\phi$ of
agents behave stochastically ($m_i^2<1$) because $Q<1$. As a consequence,
$A^\mu$ has stochastic fluctuations of order $\sqrt{N(1-Q)}$ which are of the
same order of its  average $\avg{A^\mu}\sim \sqrt{H}$. With endogenous
information, these fluctuations of
$A^\mu$ induce a  dynamics of $\mu(t)$ which is ergodic in the sense that
typically each $\mu$ is visited with a frequency $\varrho^\mu\simeq 1/P$ in the
stationary state\cite{forthcoming}. The situation changes completely when agents
follow Eq. \req{uitilde}. Indeed the system converges to a Nash equilibrium where
agents play in a deterministic way, i.e. $m_i^2=1$ (or $Q=\phi=1$). 
The noise due to the stochastic choice of $s_i$ by Eq. (\ref{sit}) is totally
suppressed. The system becomes deterministic and the dynamics of
$\mu(t)$  locks into some periodic orbit. The ergodicity assumption then
breaks  down: Only a small number $\tilde P\ll P$ of patterns $\mu$ are
visited in the stationary state of the system, whereas the others never
occur ($\varrho^\mu=0$). This leads to an effective reduction of the parameter 
$\alpha\to\tilde\alpha=\tilde P/N$, which further diminishes $\sigma^2$. 
Numerical simulations show that $\tilde P\propto \sqrt{P}$ which imply
that $\tilde\alpha\to 0$ in the limit $P=\alpha N\to\infty$, i.e
$\sigma^2/N\to 0$. 

In summary we have shown how methods of statistical physics of
disordered systems can successfully be applied to study models
of interacting heterogeneous agents. Our results extend easily to
more general models \cite{forthcoming} and, more importantly,
the key ideas can be applied to more realistic models 
of financial markets, where heterogeneities arise e.g. from asymmetric 
information.

We acknowledge J. Berg, A. De Martino, 
S. Franz, F. Ricci-Tersenghi, S. Solla, M.
Virasoro and Y.-C. Zhang for discussions and useful suggestions. 
This work was partially supported by Swiss National Science 
Foundation Grant Nr 20-46918.98.

\begin{figure}
\centerline{\psfig{file=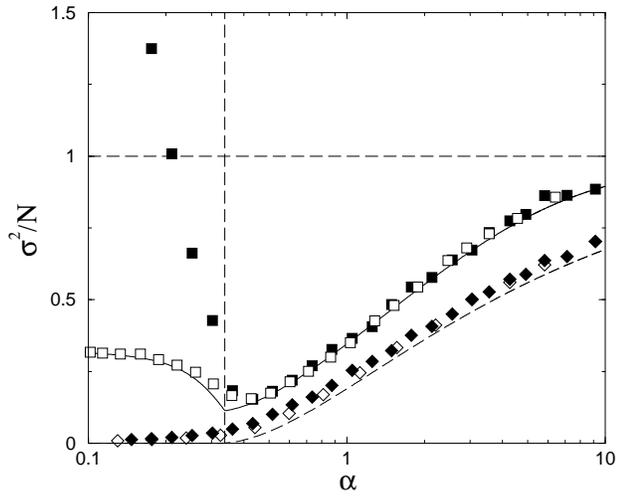,width=8cm}}
\caption{$\sigma^2/N$ versus $\alpha=P/N$ for $P=2^6$ for
inductive dynamics (full squares), for
the numerical minimization of Eq. (\ref{HID})  
(open squares), corrected
inductive  dynamics (full diamonds) and the ground state of $\sigma^2$
(open diamonds). The full and the dashed lines are the 
corresponding analytic results. Averages are taken over 200 
realizations.}
\label{fig1}
\end{figure}


\begin{thebibliography}{99}
\bibitem{nota} Both $\omega_i^\mu$ and $\xi_i^\mu$ take values
in $\{0,\pm 1\}$ but they are not independent: $\omega_i^\mu\xi_i^\mu=0$ 
and $\omega_i^\mu+\xi_i^\mu \ne 0$ for all $i,\mu$.
\bibitem{notaT} $\beta$ is introduced as a device to study the minima of 
$H$ and it should not be confused with $\Gamma$.
Eq. (\ref{dynID}) suggests that $\Gamma$ is not an inverse temperature, 
in this context, but rather the learning rate. However a system with a 
finite memory -- i.e. replacing Eq. \req{uit} by
$U_{s,i}(t+1)=(1-\epsilon)U_{s,i}(t)-a_{s,i}^{\mu(t)}G(A^{\mu(t)})$  --
is described by the properties of $H$ at a finite temperature 
$\beta =\Gamma/\epsilon$.
\end{thebibliography}
\end{document}